\documentclass[a4paper,twoside]{article}

\usepackage{soul}
\usepackage{epsfig}
\usepackage{calc}
\usepackage{amssymb}
\usepackage{amstext}
\usepackage{amsmath}
\usepackage{amsthm}

\usepackage{multicol}
\usepackage{pslatex}
\usepackage{apalike}
\usepackage{multirow}
\usepackage{booktabs}
\usepackage{algorithm2e}
\usepackage[bottom]{footmisc}
\usepackage{SCITEPRESS}     
\usepackage{xcolor}
\begin{document}

\title{Understanding Membership Inference Attacks on EEG Models}


\author{\authorname{Mehrdad Khodapanahandeh\sup{1}, Selin Unal\sup{2}, Berkay Kemal Balioglu\sup{2}, Alireza Khodaie\sup{2},\\ Beren Semiz\sup{3,4}, M. Emre Gursoy\sup{2,5}}
\affiliation{\sup{1}Department of Biomedical Sciences and Engineering, College of Engineering, Koç University, Istanbul, Türkiye}
\affiliation{\sup{2}Department of Computer Engineering, College of Engineering, Koç University, Istanbul, Türkiye}
\affiliation{\sup{3}Department of Electrical and Electronics Engineering, College of Engineering, Koç University, Istanbul, Türkiye}
\affiliation{\sup{4}Koç University Research Center for Translational Medicine (KUTTAM), Koç University, Istanbul, Türkiye}
\affiliation{\sup{5}Koç University \& Iş Bank Artificial Intelligence Center (KUIS AI), Koç University, Istanbul, Türkiye}
\email{\{mkhodapanahandeh21, sunal22, bbalioglu23, akhodaie22, besemiz, emregursoy\}@ku.edu.tr}
}


\abstract{Electroencephalography (EEG) signals contain sensitive, subject-specific information, making models trained on them potential targets for membership inference attacks (MIAs). Although MIAs have been studied extensively in other domains, their applications to EEG models are relatively scarce, and the factors driving the vulnerability of EEG models to MIAs are not adequately understood. In this paper, we present a systematic empirical analysis of MIAs on EEG models using three datasets, seven model architectures, four attack methods, and two metrics. Results show that poor model generalization (large train-test loss and accuracy gap) is the strongest indicator of MIA vulnerability. Longer training and larger models also increase vulnerability. Moreover, no particular scalp region, frequency band, inter-channel correlation pattern, or temporal versus spectral representation of EEG data consistently explains MIA vulnerability across datasets. Finally, subject identification and leave-one-subject-out experiments indicate that models relying more heavily on subject-specific representations are more vulnerable to MIAs. Overall, our findings identify the key drivers of membership leakage in EEG models and can provide guidance for developing MIA-resilient EEG systems.}

\onecolumn \maketitle \normalsize \setcounter{footnote}{0} \vfill

\section{\uppercase{Introduction}}
\label{sec:introduction}

Electroencephalography (EEG) has become one of the most widely used modalities for non-invasive monitoring of brain activity. It underpins applications ranging from clinical diagnosis of epilepsy and neurological disorders to brain-computer interfaces (BCIs) and increasingly to consumer devices for gaming, meditation, and cognitive monitoring \cite{liu2025recent,portillo2021mind}. The proliferation of low-cost consumer EEG headsets has moved this technology out of the laboratory and into everyday settings, and enabled large-scale collection of EEG data \cite{jwa2022addressing,yang2025regulating}.

On the other hand, this shift raises significant privacy concerns, since prior works have shown that EEG signals can be used to perform a variety of inference attacks, e.g., identity, sex, age, emotional and cognitive state, and presence of medical conditions \cite{meng2024protecting,mandal2022sok}. One type of inference attack is the \textit{membership inference attack (MIA)}, which enables an attacker to determine whether a particular subject's data was used to train a deep learning model \cite{shokri2017membership}. MIAs are a significant threat to EEG models; for example, inferring that an individual's EEG was used to train an epilepsy or depression screening model can reveal that this individual has epilepsy or depression. This makes it important to understand when and why EEG models are vulnerable to MIAs, and which factors drive MIA vulnerability in EEG models.

In this paper, we conduct a systematic empirical investigation into the factors that influence the vulnerability of EEG models to MIAs. To the best of our knowledge, the pioneering study in this field is the work of \cite{cobilean2025investigating}, which provides initial evidence that CNN-based EEG models are susceptible to the loss attack. We use this work as our starting point and broaden the evaluation by considering three EEG datasets representing different acquisition settings and classification tasks, six CNN architectures, a controllable CNN architecture (TestNetwork), and four MIA methods. 

Beyond measuring attack success, our central objective is to explain \textit{when} and \textit{why} membership leakage occurs in EEG models. In that regard, our main findings and contributions are:
\begin{itemize}
    \item We perform a comprehensive evaluation of MIAs against EEG models using MoBI, Driver-Drowsiness, and Alzheimer's Disease datasets. We evaluate the loss, confidence, modified entropy, and offline LiRA attacks using AUC and TPR at 1\% FPR metrics. While the loss, confidence, and modified entropy attacks perform similarly, LiRA yields greater leakage.
    \item We show that model generalization is strongly associated with MIA vulnerability. Larger train-test loss and accuracy gaps consistently correspond to higher attack AUCs. Also, increasing the model training duration increases both the generalization gap and MIA vulnerability, highlighting overfitting as a central driver of MIA vulnerability.
    \item Our experiments which disentangle the number of participants and the amount of EEG data contributed by each participant show that increasing the number of distinct participants generally reduces attack success, whereas varying the amount of EEG data per participant has a comparatively weaker effect. 
    \item We examine the effects of model size and EEG signal characteristics. MIA vulnerability generally increases with model size, although it begins to saturate for larger models. In contrast, no individual scalp region, frequency band, inter-channel correlation pattern, or temporal versus spectral representation consistently explains membership leakage across EEG datasets.
    \item We investigate the role of subject-specificity through subject identification and leave-one-subject-out experiments. We find that models whose task performance deteriorates more when evaluated on unseen subjects also exhibit greater MIA vulnerability, suggesting that reliance on subject-specific representations is an important source of membership leakage in EEG models.
\end{itemize}

Overall, our findings indicate that MIA vulnerability in EEG models is primarily associated with poor generalization, overfitting, and reliance on subject-specific representations, rather than with any single spatial or spectral component of the EEG signal. These results provide practical guidance for evaluating privacy risks and designing EEG systems that are less susceptible to MIAs.

\section{\uppercase{Related Work}} \label{sec:RelatedWork}

MIAs were introduced by \cite{shokri2017membership}, who showed that an attacker with black-box access to a classification model can determine whether a given sample is part of the model's training set. Subsequent work by \cite{yeom2018privacy} established the connection between overfitting and MIA vulnerability, and proposed the \textbf{loss} attack for membership inference. \cite{song2021systematic} argued that aggregate notions such as attack accuracy can understate MIA risk, and introduced metric-based attacks. \cite{carlini2022membership} proposed that MIAs should be assessed by their true positive rates (TPR) at low false positive rates (FPR), rather than average-case accuracy or area under curve (AUC). They proposed the Likelihood Ratio Attack (LiRA), which estimates the distribution of model outputs when the sample is included versus excluded from training, and scores membership by a per-sample likelihood ratio. \cite{hu2022membership} and \cite{bai2024membership} provide surveys that cover MIAs in centralized and federated environments.

While MIAs have been studied extensively for image, tabular, and textual data, they have only recently been considered in the context of EEG. To the best of our knowledge, \cite{cobilean2025investigating} presented the first investigation of MIAs against EEG models, reporting that EEG models are vulnerable to the loss attack. We consider \cite{cobilean2025investigating} as the starting point, but create a broader setup with multiple datasets and MIA methods. Recent work by \cite{khanam2026evaluating} also highlights membership inference risk in federated EEG learning.

Beyond MIAs, concerns about privacy in EEG systems have been reported in several works. Studies have shown that EEG recordings passively encode a range of attributes that the user did not intend to share, including sex, age, and cognitive or emotional state \cite{meng2024protecting,mandal2022sok}. While these attacks are relevant, they differ from MIAs, which are the main focus of our paper. The subject-specific nature of EEG (i.e., EEG signals differ from individual to individual) has also been reported in several studies, such as \cite{campisi2014brain,demuru2020eeg}. This motivates our analysis of the connections between MIA vulnerability and subject-specificity in Section \ref{sec:subjID}.


\vspace{-8pt}

\section{\uppercase{Methodology}} \label{sec:Methodology}

\vspace{-6pt}
\subsection{Problem Setting}

Let $\mathcal{D}_{\mathrm{train}}=\{(x_i,y_i)\}_{i=1}^{N}$ denote the training dataset consisting of EEG segments (we use \textit{segment} and \textit{sample} interchangeably), where $x_i\in\mathbb{R}^{C\times T}$ represents an EEG segment with $C$ channels and $T$ timestamps, and $y_i\in\{1,\ldots,K\}$ denotes its corresponding class label. A classification model $f_\theta$, parameterized by $\theta$, is trained on $\mathcal{D}_{\mathrm{train}}$ to predict the class labels of unseen test samples. We denote the loss of the model on sample $(x,y)$ by $\ell_\theta(x,y)$. For the classification task, given an input $x$, the model produces a logit vector:
\begin{equation}
    \mathbf{z}=f_\theta(x)\in\mathbb{R}^{K}
\end{equation}
The logits can be converted into class probabilities using the softmax function:
\begin{equation}
    p_{\theta,k}(x)
    =\operatorname{softmax}(\mathbf{z})_k
    =\frac{\exp(z_k)}
    {\sum_{j=1}^{K}\exp(z_j)}
\end{equation}
The predicted class is consequently given by:
\begin{equation}
    \hat{y}_\theta(x)
    =\underset{k\in\{1,\ldots,K\}}{\arg\max}\;
    p_{\theta,k}(x)
\end{equation}

\subsection{Datasets and Preprocessing}

We use three datasets in our setup: MoBI, Driver-Drowsiness, and Alzheimer's Disease. Below, we briefly explain how the datasets were obtained and preprocessed before model training. 

\textbf{Multi-Modal Mobile Brain-Body Imaging (MoBI) Dataset} comprises EEG and head movement recordings collected from children during eyes-open rest and creative video game play \cite{ravindran2017multi}. EEG was acquired at 220 Hz using a four-channel consumer-grade dry-electrode Muse headset (AF7, AF8, TP9, TP10; reference near FPz), with a per-channel 10 Hz headband status signal used for electrode contact quality control. Following \cite{cobilean2025investigating}, we use the 11 participants with both rest and video game recordings, and train the model $f_\theta$ to separate EEG segments recorded during rest versus video game play. 

Our preprocessing of the dataset follows the original dataset methodology \cite{sujatha2019assaying} and is applied separately to rest and game segments. Each EEG channel is band-pass filtered at 1-35 Hz using a fourth-order zero-phase Butterworth filter, downsampled from 220 to 110 Hz, cleaned using Artifact Subspace Reconstruction (ASR) \cite{mullen2015real} via MEEGkit, followed by ±4-SD z-score clipping and normalization. Model inputs are formed using 2.5 second windows, with overlap varied systematically: 0\%, 25\%, 50\%, 75\%, and approximately 99.6\% (1 sample stride), the latter matching prior work \cite{cobilean2025investigating}. Windows are excluded when the headband status signal indicates poor electrode contact (status $>$ 3) for more than 30\% of their duration. 

\textbf{Driver-Drowsiness Dataset} contains drivers' EEG recordings acquired during a simulated driving setup in which participants were required to maintain a straight driving trajectory while random lane deviation events were introduced \cite{cao2019multi}. The recordings were subsequently labeled for alertness and drowsiness in later work \cite{cui2022compact}. The original recordings contain 30-channel EEG acquired at 500 Hz. We preprocessed the EEG recordings by filtering between 1 and 50 Hz and downsampled to 128 Hz. Following \cite{cui2022compact}, 3 second EEG segments preceding deviation events were extracted and labeled as alert or drowsy based on reaction time criteria described in prior work. The data were then balanced across classes, resulting in 2022 segments (1011 alert + 1011 drowsy). Each sample contains 384 time points per channel across 30 EEG channels. The model is trained to identify whether each segment corresponds to an alert or drowsy state.

\textbf{Alzheimer's Disease Dataset} contains rest-state, eyes-closed EEG recordings from 88 participants \cite{miltiadous2023dataset}. We used EEG data from 36 patients with Alzheimer’s disease and 29 healthy control subjects (cognitively normal). EEG was recorded from 19 scalp channels positioned according to the international 10-20 system at a sampling rate of 500 Hz. The original EEG recordings were acquired with a 0.5-45 Hz band-pass filter. For our experiments, the signals were resampled to 160 Hz and segmented into 1 second windows, yielding 160 time points per channel. These fixed-length segments were used as inputs for the EEG classification and membership inference experiments. The classification model is trained to discriminate between EEG segments from Alzheimer's disease versus cognitively normal.

\subsection{Model Architectures}

Following the prior work of \cite{cobilean2025investigating}, we use six convolutional neural network (CNN) architectures as our target models. In addition, for our model complexity analysis (Section \ref{sec:modelchar}), we introduce a controllable CNN architecture named TestNetwork. All models have on the order of $10^4$ parameters in their default configuration, consistent with the compact model sizes used in prior work. We train the models with binary cross-entropy loss, Adam optimizer, learning rate $1 \times 10^{-3}$, batch size 128, and 70\%-15\%-15\% train-validation-test splits.

\textit{ShallowConvNet} is a shallow neural network with two convolutional layers, followed by a ReLU activation, a mean pooling layer, and a fully connected output layer. Its first convolutional layer performs temporal feature extraction, and its second convolutional layer is a spatial layer that learns the dependencies between EEG channels \cite{schirrmeister2017deep}.

\textit{EEGNet} is a deeper counterpart of ShallowConvNet. Its first layer performs standard 2D convolution followed by a depthwise convolutional layer, which is then followed by a pointwise convolution to combine outputs across channels \cite{lawhern2018eegnet}.

\textit{EEGNet2TL} is a modified version of EEGNet by integrating an additional layer into the temporal feature extractor, for extracting temporal features using different resolutions \cite{cobilean2025investigating}.

\textit{EEGInception} adapts the inception module to EEG classification. It consists of two inception blocks and each block consists of branches that extract temporal features using kernels with different lengths. Outputs from all branches are concatenated, passed through average pooling, and convolutions are performed before passing them to the final classification layer \cite{santamaria2020eeg}.

\textit{TIDNet} consists of two stages: temporal feature extraction and spatial feature extraction. They are fed into a classification layer, which is a fully connected layer \cite{kostas2020thinker}.

\textit{Conv1DModel} is a lightweight one-dimensional CNN. It consists of three 1D convolutional layers. The first one uses a kernel size of 64, and each subsequent layer has a kernel size that is half of the previous one \cite{kiranyaz20211d}. 

\textit{TestNetwork} contains a stack of temporal convolutional layers followed by a spatial convolution and a linear classifier. The number of temporal filters per layer and the number of layers can be varied independently, allowing us to change the model size over a wide range ($\sim$1.9K to $\sim$830K parameters) while holding the rest of the architecture fixed. Our default TestNetwork is the smallest one, containing 1 temporal layer and 20 filters per layer.

\subsection{MIAs and Attack Methods} \label{sec:MIAattacks}

Given an EEG sample $(x,y)$ and black-box access to the trained classification model $\theta$, the objective of a membership inference attack (MIA) is to determine whether $(x,y) \in \mathcal{D}_{\mathrm{train}}$, i.e., whether the sample was used to train the target model. There are multiple ways in which a MIA can be conducted. The original work by \cite{cobilean2025investigating} uses the \textbf{loss} attack. In this paper, we use the following four attacks. 

\textbf{Loss attack} is based on the intuition that $\theta$ would typically produce lower loss on member samples than non-member samples, since it is trained with the goal of minimizing loss on $\mathcal{D}_{\mathrm{train}}$ \cite{yeom2018privacy}. The attack computes $\ell_\theta(x,y)$ and compares it with a decision threshold. Samples with lower loss are predicted as members.

\textbf{Confidence attack} is based on the intuition that $\theta$ classifies member samples more confidently than non-members, since it has seen those members during training. The attack computes the maximum predicted class probability:
\begin{equation}
c_\theta(x)=\max_{k\in\{1,\ldots,K\}}p_{\theta,k}(x) = p_{\theta,\hat{y}_\theta(x)}(x)
\end{equation}
and compares it with a decision threshold. It predicts $(x,y)$ as a member if $c_\theta(x)\geq$ that threshold, and as a non-member otherwise.

\textbf{Modified Entropy attack} measures the uncertainty of the model's prediction while accounting for the ground-truth label $y$ \cite{song2021systematic}. For a given sample $(x,y)$, its modified prediction entropy is computed as:
\begin{equation}
\begin{split}
\operatorname{Mentr}_\theta(x,y)
    ={}&-\bigl(1-p_{\theta,y}(x)\bigr)\log p_{\theta,y}(x)\\
    &-\sum_{\substack{k=1\\k\neq y}}^{K}
    p_{\theta,k}(x)\log\bigl(1-p_{\theta,k}(x)\bigr)
\end{split}
\end{equation}
This score decreases as the probability assigned to the correct class increases, and increases as the probabilities assigned to incorrect classes increase. Since member samples are expected to produce lower modified entropy, the attack predicts $(x,y)$ as a member if $\operatorname{Mentr}_\theta(x,y)\leq\tau_y$, where $\tau_y$ is a class-specific decision threshold, and as a non-member otherwise.

\textbf{Likelihood Ratio (LiRA) attack} estimates how models that have not been trained on a candidate sample $(x,y)$ behave on that sample \cite{carlini2022membership} (this is the ``offline'' variant of LiRA). The attacker trains multiple shadow models (we train 64 shadow models) on auxiliary datasets that do not contain $(x,y)$ and computes the logit-scaled confidences:
\begin{equation}
    \phi_\theta(x,y)
    =\log\left(
    \frac{p_{\theta,y}(x)}
    {1-p_{\theta,y}(x)}
    \right)
\end{equation} 
These are used to estimate the sample-specific OUT distribution
$\mathcal{N}\bigl(\mu_{\mathrm{out}}(x,y),
\sigma_{\mathrm{out}}^2(x,y)\bigr)$, where $\mathcal{N}(\cdot)$ denotes the normal distribution. Letting $\Phi(\cdot)$ denote the cumulative distribution function of the standard normal distribution, the LiRA score is:
\begin{equation}
    \operatorname{LiRA}_\theta(x,y)
    =
    \Phi\left(
    \frac{\phi_\theta(x,y)-\mu_{\mathrm{out}}(x,y)}
    {\sigma_{\mathrm{out}}(x,y)}
    \right)
\end{equation}
A high score indicates that the target model assigns an unusually high confidence to the correct class compared with models that did not observe $(x,y)$. The attack therefore predicts $(x,y)$ as a member if $\operatorname{LiRA}_\theta(x,y)
\geq\tau_{\mathrm{LiRA}}$, and non-member otherwise.

\subsection{Evaluation Metrics}

We evaluate MIAs using a balanced set of members and non-members. A true positive occurs when a member sample is predicted as a member, a false positive occurs when a non-member is predicted as a member, a true negative is when a non-member is predicted as a non-member, and a false negative is when a member is predicted as a non-member. Then, we use two metrics to measure MIA effectiveness. 

\textbf{Attack AUC:} We vary the membership decision thresholds of MIAs to obtain the receiver operating characteristic (ROC) curve. AUC corresponds to the area under the ROC curve. AUC = 0.5 is equivalent to random guess (no leakage) and AUC = 1.0 is a perfect attack. We use AUC as our primary metric because of its popularity and its nature as an average-case metric, i.e., it rewards separating members and non-members generally, rather than focusing on edge cases.

\textbf{TPR @ 1\% FPR:} Following \cite{carlini2022membership}, we report the true positive rate (the fraction of members correctly identified) at a fixed false positive rate (FPR) of 1\%. This is an edge-case metric: it quantifies how many members an attacker can find, while almost never falsely predicting a non-member as a member.


\section{\uppercase{Analysis and Discussion}} \label{sec:Analysis}

In this section, we present our empirical analysis regarding the key factors influencing MIA vulnerability in EEG models. To quantify how strongly a factor relates to MIA vulnerability, we often report the Pearson correlation coefficient $r$ between that factor and the attack success (measured primarily using the AUC metric), as well as the p-value for statistical significance, in addition to relevant visuals and plots. The coefficient $r\in[-1,1]$ measures the strength and direction of linear correlation between a factor and MIA vulnerability. Values near +1 or -1 indicate significant correlation, whereas values near 0 indicate no relationship. For p-values, we treat $p < 0.05$ as statistically significant. 

\subsection{Model Generalization}

We find that a key factor affecting MIA vulnerability is how well the model generalizes (versus overfits) to the training dataset. To quantify this, we use two notions: the train-test \textbf{loss gap} and \textbf{accuracy gap}. Let $\mathcal{D}_{\mathrm{test}}=\{(x_j,y_j)\}_{j=1}^{M}$ denote the test dataset disjoint from $\mathcal{D}_{\mathrm{train}}=\{(x_i,y_i)\}_{i=1}^{N}$. The average training and test losses are:
\begin{equation}
    L_{\mathrm{train}}
    =\frac{1}{N}\sum_{i=1}^{N}\ell_\theta(x_i,y_i)
    ~~~~
    L_{\mathrm{test}}
    =\frac{1}{M}\sum_{j=1}^{M}\ell_\theta(x_j,y_j)
\end{equation}
The train-test loss gap is then defined as:
\begin{equation}
    G_{\mathrm{loss}}
    =L_{\mathrm{test}}-L_{\mathrm{train}}
\end{equation}
A large positive $G_{\mathrm{loss}}$ indicates that the model produces considerably lower loss on its training samples than on unseen samples, suggesting stronger overfitting. Similarly, the training and test accuracies are defined as:
\begin{equation}
\begin{split}
A_{\mathrm{train}}
    =\frac{1}{N}\sum_{i=1}^{N}
    \mathbb{I}\!\left[\hat{y}_\theta(x_i)=y_i\right] \\
    A_{\mathrm{test}}
    =\frac{1}{M}\sum_{j=1}^{M}
    \mathbb{I}\!\left[\hat{y}_\theta(x_j)=y_j\right]
\end{split}
\end{equation}
where $\mathbb{I}[\cdot]$ is the indicator function. The train-test accuracy gap is defined as:
\begin{equation}
    G_{\mathrm{acc}}
    =A_{\mathrm{train}}-A_{\mathrm{test}}
\end{equation}
Thus, larger positive values of both $G_{\mathrm{loss}}$ and $G_{\mathrm{acc}}$ indicate poorer generalization. 

\begin{figure*}[!t]
  \centering
   {\epsfig{file = 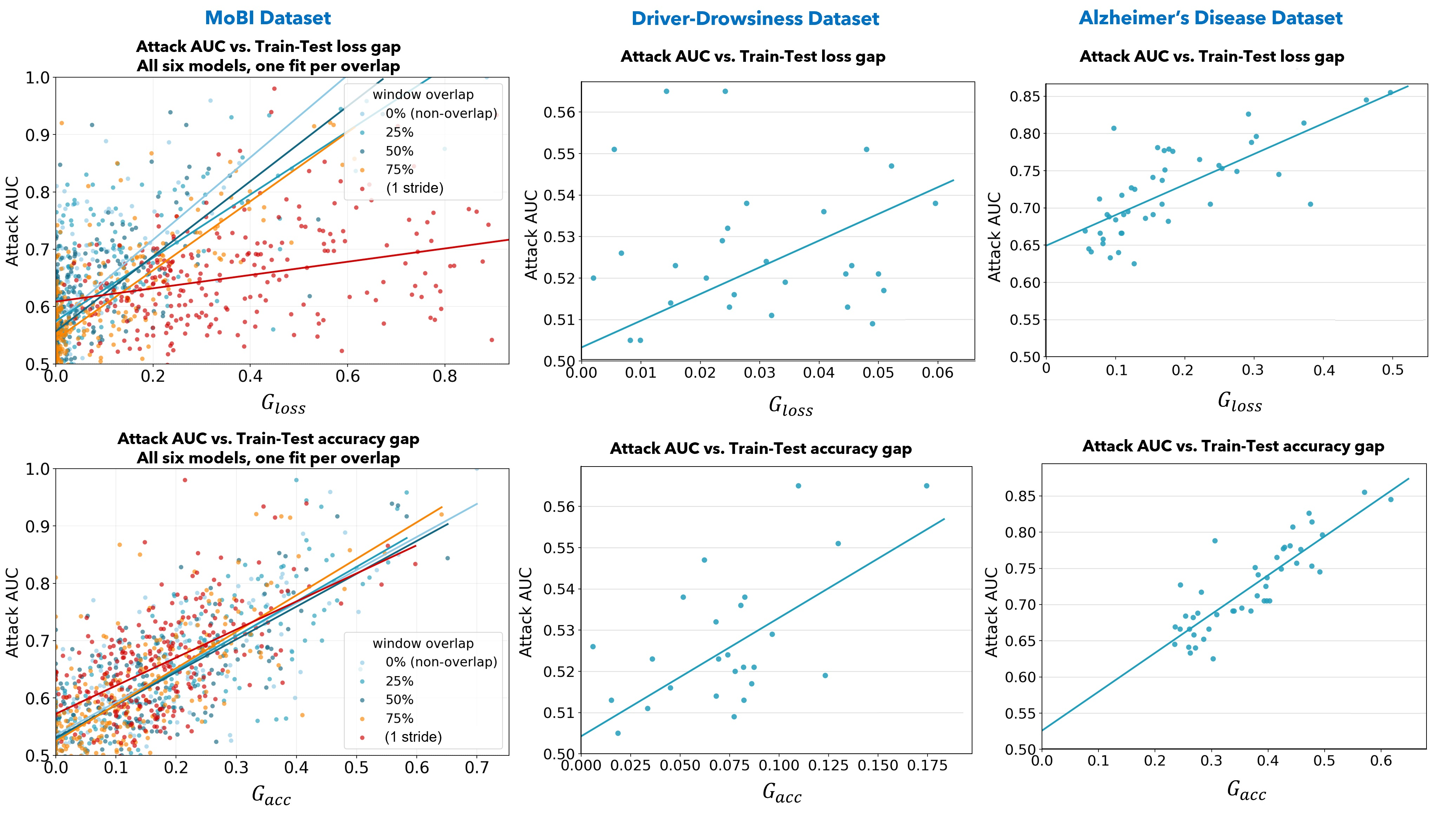, width=.95\textwidth}}
  \caption{\textbf{Model generalization results.} First row: attack AUC vs loss gap, second row: attack AUC vs accuracy gap. Left pair is for the MoBI dataset at five overlaps of EEG windowing, middle pair for the Driver-Drowsiness dataset, and right pair for the Alzheimer's Disease dataset (latter two are with non-overlapping windowing). }
  \label{fig:gapvsauc}
\end{figure*}

Results relating attack AUC to $G_{\mathrm{loss}}$ and $G_{\mathrm{acc}}$ are given in Figure \ref{fig:gapvsauc}. The results show a positive and highly significant correlation between attack AUC versus $G_{\mathrm{loss}}$ and $G_{\mathrm{acc}}$ (Pearson $r$ between 0.59 and 0.80, p-values below 0.001). In other words, when the performance of the model is different on the training dataset versus the test dataset (e.g., the model does not generalize well), attack AUCs increase. According to the MoBI dataset, the relationship is somewhat weaker with 1-stride EEG windowing than with other (more conventional) overlaps, but the remaining datasets with no window overlap also show a significant positive correlation (Pearson $r$ between 0.56 and 0.68 for Driver-Drowsiness; between 0.74 and 0.85 for Alzheimer's Disease dataset).

\subsection{Dataset Characteristics}

\begin{figure*}[!t]
  \centering
   {\epsfig{file = 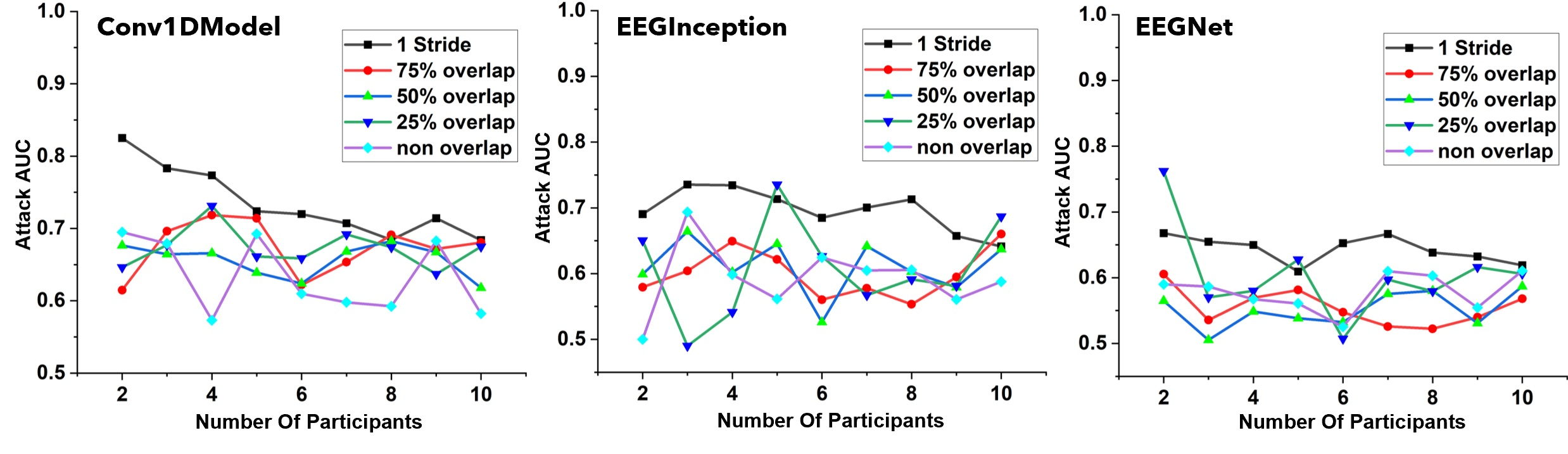, width=.9\textwidth}}
  \caption{\textbf{Attack AUC vs number of participants.} Increasing the number of participants reduces the attacker's AUC, despite some fluctuations (MoBI dataset).}
  \label{fig:aucvsparticipant}
\end{figure*}

\begin{figure}[t]
  \centering
   {\epsfig{file = 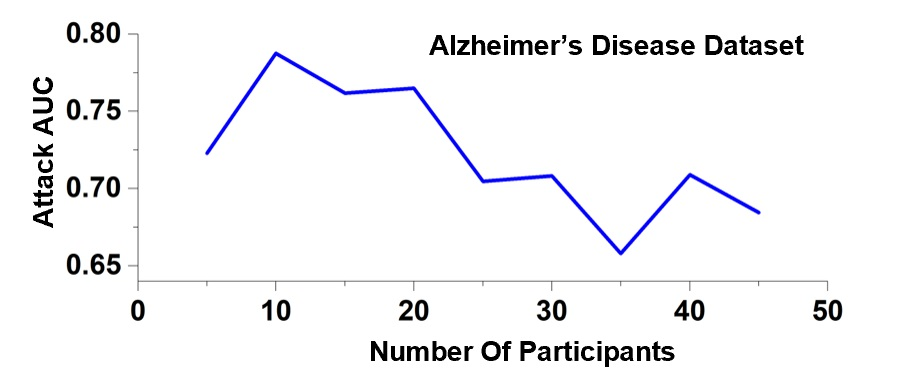, width=.9\columnwidth}}
  \caption{\textbf{Attack AUC vs number of participants} for the Alzheimer's Disease dataset with non-overlap windowing. The results are similar to the MOBI dataset, with fluctuations but decreasing overall.}
  \label{fig:aucvsparticipantalzanddriv}
\end{figure}

Next, we study whether different characteristics of EEG datasets affect MIA vulnerability. In particular, we train the models after limiting how many participants' data is included in $\mathcal{D}_{\mathrm{train}}$, e.g., using only 1 participant, 2 participants, 3 participants, and so forth. The resulting attack AUCs are shown in Figure \ref{fig:aucvsparticipant}. We observe from the figure that attack AUCs generally fall as the number of participants increases. This trend seems to hold on multiple models (Conv1DModel, EEGInception, EEGNet). As illustrated in Figure \ref{fig:aucvsparticipantalzanddriv}, the AUCs of the Alzheimer's Disease dataset are consistent with the trend of the MoBI dataset with non-overlap windowing.

\begin{figure}[!t]
  \centering
   {\epsfig{file = 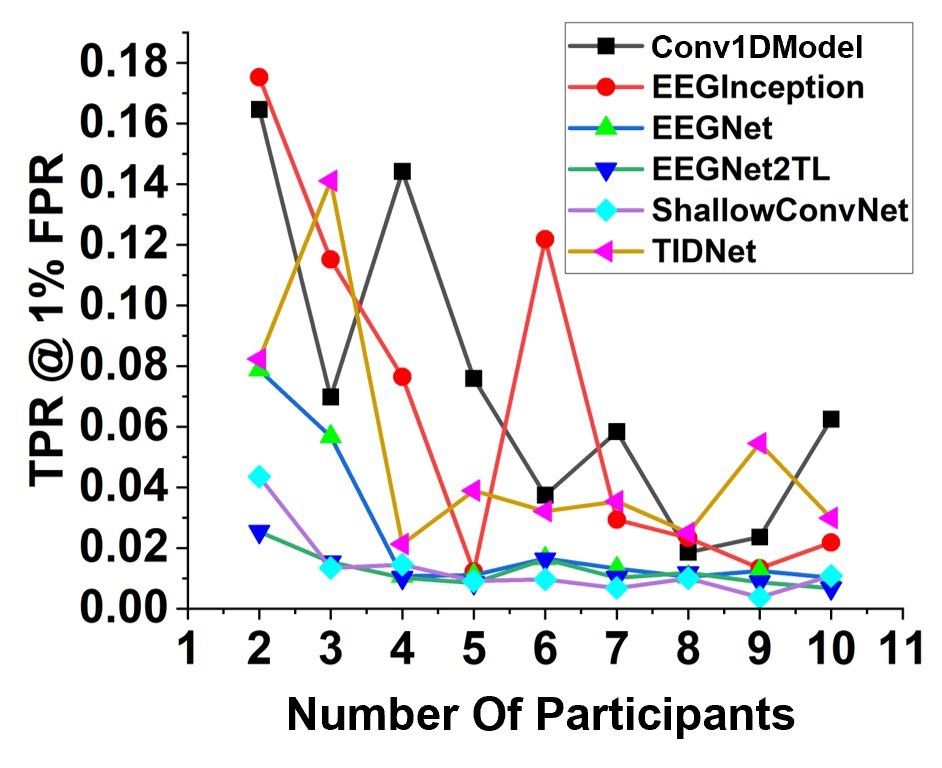, width=.7\columnwidth}}
  \caption{\textbf{TPR @ 1\% FPR vs number of the participants.} This experiment is done just for 1 stride window overlap of the MoBI dataset.}
  \label{fig:tprvsparticipant}
\end{figure}

While the above finding is consistent with the prior work of \cite{cobilean2025investigating}, the TPR @ 1\% FPR metric behaves differently. As shown in Figure \ref{fig:tprvsparticipant}, TPRs decrease with increasing number of participants (from 0.18 to as low as 0.006). The correlation between TPR @ 1\% FPR and the number of participants is negative and statistically significant (according to p-value) across all six models.

\begin{figure*}[t]
  \centering
   {\epsfig{file = 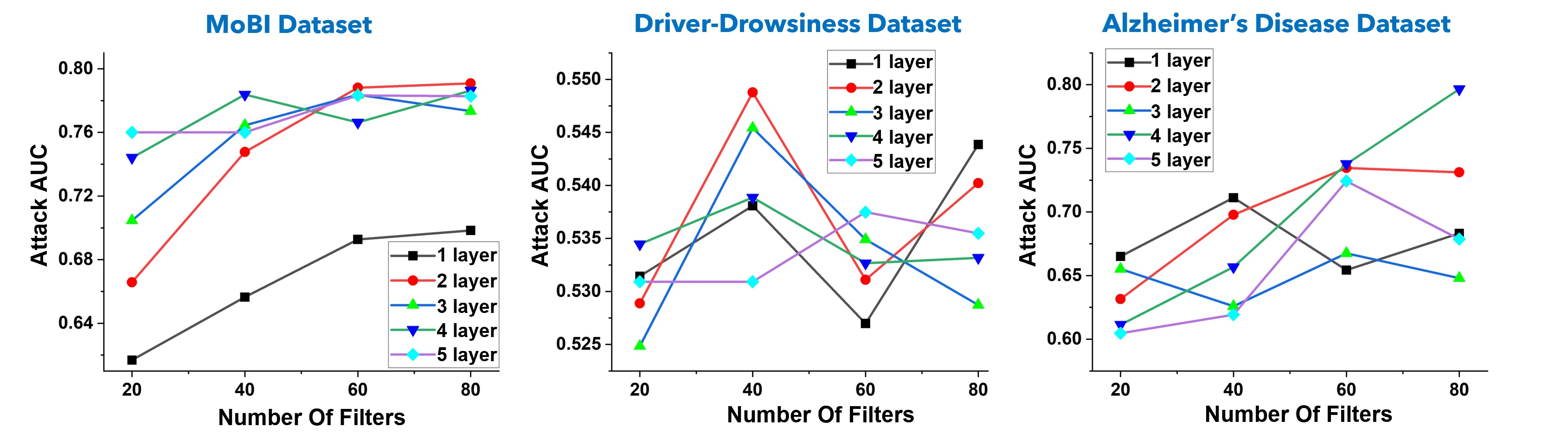, width=.9\textwidth}}
  \caption{\textbf{Impact of model size (filter and layer count).} MoBI dataset (left), Driver-Drowsiness dataset (middle), and Alzheimer's disease dataset (right) with non-overlap windowing.}
  \label{fig:filtervsauc}
\end{figure*}

\begin{figure}[t]
  \centering
   {\epsfig{file = 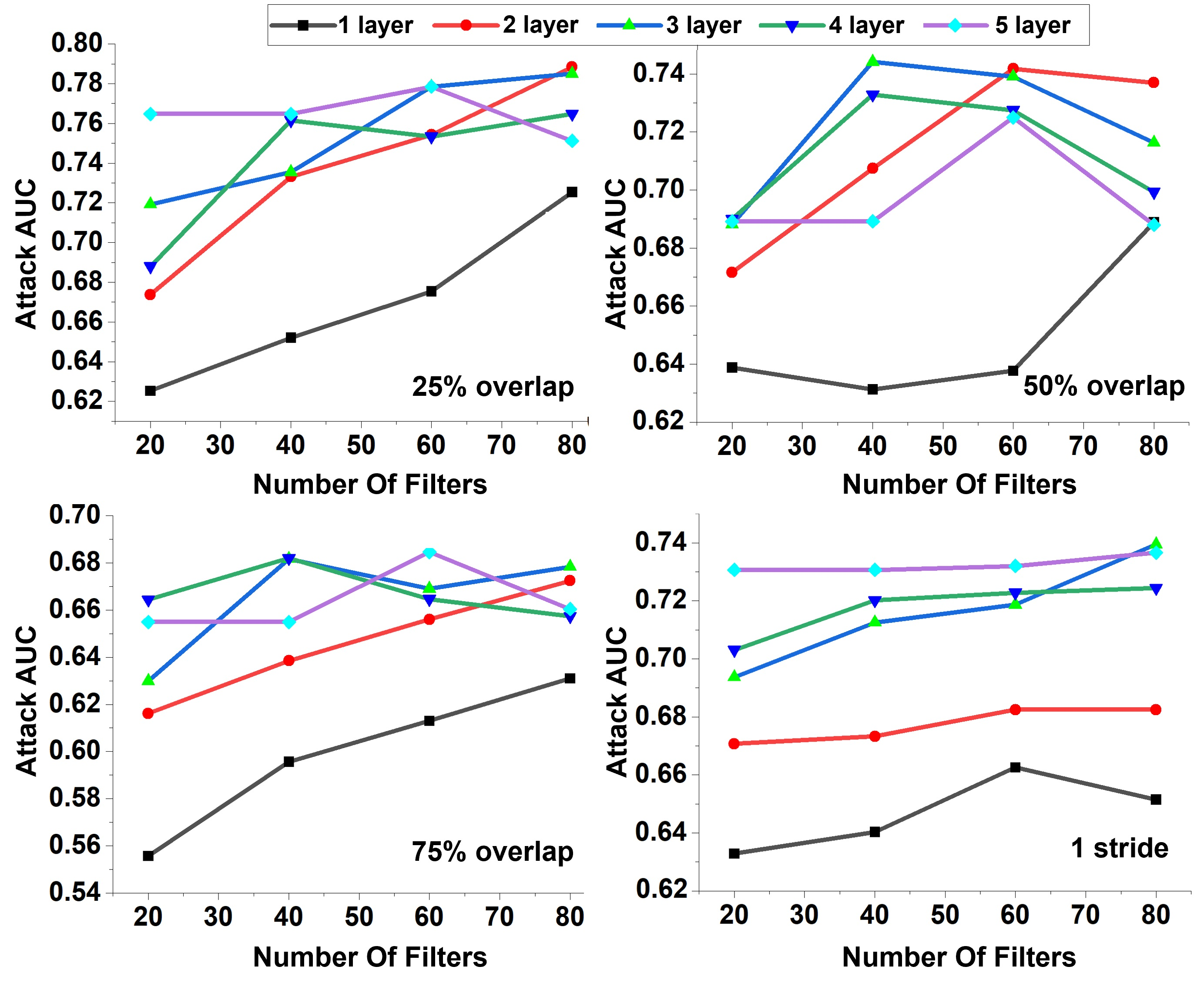, width=\columnwidth}}
  \caption{Attack AUC versus number of filters per layer for the controllable TestNetwork, for 25\%, 50\%, 75\%, and 1 stride window overlap (MoBI dataset).}
  \label{fig:filtervsauc2}
\end{figure}

We also perform an experiment in which we hold the number of participants fixed and vary only the amount of training windows, i.e., how much data from each participant is included in $\mathcal{D}_{\mathrm{train}}$. We do not have any plots regarding this experiment due to the page limit, but we discuss the results verbally. We find that the training dataset size has a weak effect on MIA vulnerability. Across 30 model and EEG window overlap conditions, the mean correlation between the training dataset size and attack AUC was -0.16, and for TPR @ 1\% FPR the mean correlation was -0.11. These are neither strong correlations nor statistically significant according to p-value. We highlight this as a notable finding because in the prior work of \cite{cobilean2025investigating}, training dataset size and number of participants were confounded. When the two are separated as in our experiment, we find that training dataset size matters relatively little, but number of participants matters more significantly.

Overall, these results indicate that the MIA vulnerability is driven by the number of distinct individuals. More participants in the training makes each participant less memorable, i.e., less susceptible to MIA. In contrast, smaller or larger volume of data per participant does not impact MIAs as significantly. 

\subsection{Model and Training Characteristics} \label{sec:modelchar}

In this section, we study whether the characteristics of the EEG models and their training (e.g., size and complexity of the CNNs) impact MIA vulnerability, as it is often believed that larger models are more prone to overfitting and memorization. First, using our controllable TestNetwork, we vary the number of layers $\in\{1, 2, 3, 4, 5\}$ and the filters per layer $\in\{20,40,60,80\}$. The results of this experiment are shown in Figure \ref{fig:filtervsauc}. We observe that, in general, attack AUC tends to increase with number of layers and filters. The correlation between the number of parameters in the resulting CNN and attack AUC is positive and significant ($r=+0.34$ $p < 0.001$). Nevertheless, the effect saturates beyond roughly $10^5$ parameters. These findings are consistent with the intuition that larger models with a higher number of parameters are more vulnerable to memorization and MIAs. We also note that the effect is clearer on the MoBI and Alzheimer's Disease datasets, whereas it is relatively weaker on the Driver-Drowsiness dataset, indicating that the dataset can also play a role.

In addition, we perform the experiment in Figure \ref{fig:filtervsauc2} to analyze whether the window overlap amounts in MoBI dataset preprocessing affect the aforementioned findings. We report results with 25\%, 50\%, 75\%, and 1-stride window overlap. In general, results support the previous finding that increasing the number of filters increases attack AUC. Nevertheless, the 1-stride overlap experiment shows the weakest correlation because consecutive windows have large overlapping sections. Furthermore, increasing the number of layers from 1 to 2 seems to have a significant impact, whereas the remaining layer counts (2, 3, 4, 5) can often have similar AUCs.

\begin{figure*}[t]
  \centering
   {\epsfig{file = 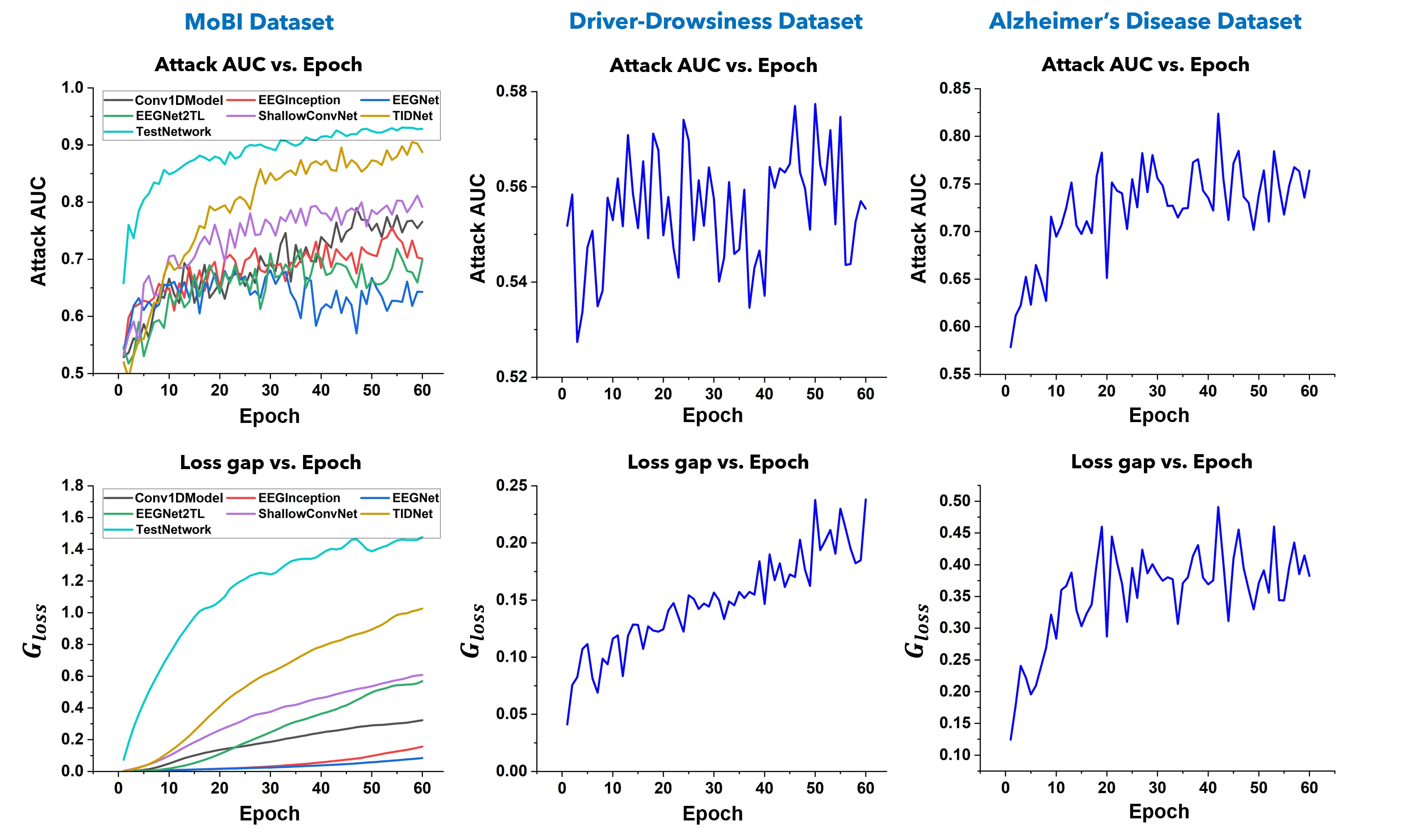, width=.9\textwidth}}
  \caption{\textbf{Number of epochs used in model training vs attack AUC and train-test loss gap.} From left to right: MoBI, Driver Drowsiness, and Alzheimer's Disease datasets. Results show that as the number of epochs increases, attack AUC and train-test loss gap increase.}
  \label{fig:epochvsaucandoverfitting}
\end{figure*}

Next, we analyze the impact of the number of epochs (i.e., training duration) on attack AUC and the train-test loss gap $G_{\mathrm{loss}}$. The results are given in Figure \ref{fig:epochvsaucandoverfitting}. We observe that attack AUC increases with the number of epochs (per-model $r$ from +0.24 to +0.75, all statistically significant according to p-value). We also observe that $G_{\mathrm{loss}}$ increases, meaning that the loss difference between training and test datasets widens. It is intuitive that the number of epochs is positively correlated with both AUC and $G_{\mathrm{loss}}$: as the model is trained more and more on $\mathcal{D}_{\mathrm{train}}$, it starts overfitting to $\mathcal{D}_{\mathrm{train}}$, widening the loss gap and increasing MIA vulnerability.

\subsection{Spatial and Spectral Predictors of EEG Signals}

Next, we examine whether MIA vulnerability can be attributed to a specific spatial, spectral, or representational property of the EEG signal. Across these analyses, no single data characteristic consistently emerged as a dominant predictor of attack success. We detail our analyses below. 

\begin{figure*}[t]
  \centering
   {\epsfig{file = 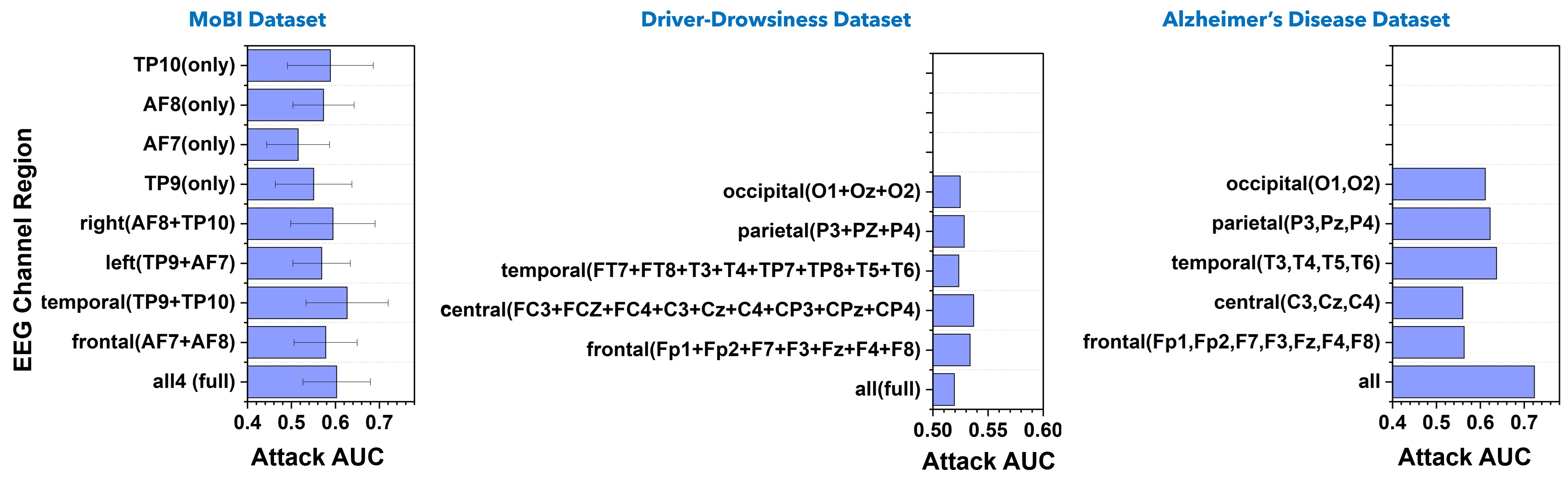, width=.9\textwidth}}
  \caption{\textbf{EEG channel region.} Attack AUC by channel subset for three datasets. No single region consistently dominates; the full montage performs comparably to or better than any subset.}
  \label{fig:regionvsauc}
\end{figure*}

\textbf{EEG Channel Regions:} Figure \ref{fig:regionvsauc} compares attack AUC across different EEG channel subsets in the three datasets. In MoBI, frontal (AF7-AF8), temporal (TP9-TP10), hemispheric, and individual-channel configurations yield broadly similar attack performance, with AUCs generally ranging from approximately 0.55 to 0.65. Similarly, the Driver-Drowsiness dataset shows little regional differentiation, with all channel subsets producing AUCs within a narrow range of approximately 0.50-0.60. Larger regional differences are observed in the Alzheimer’s Disease dataset, where the full montage yields the highest attack AUC and occipital channels show lower AUCs than the other channel groups. However, the regional pattern is not consistent across datasets. Overall, these findings indicate that membership vulnerability is not localized to a specific scalp region, and using the full set of available channels can yield higher attack AUCs than some restricted subsets.

\begin{figure}[t]
  \centering
   {\epsfig{file = 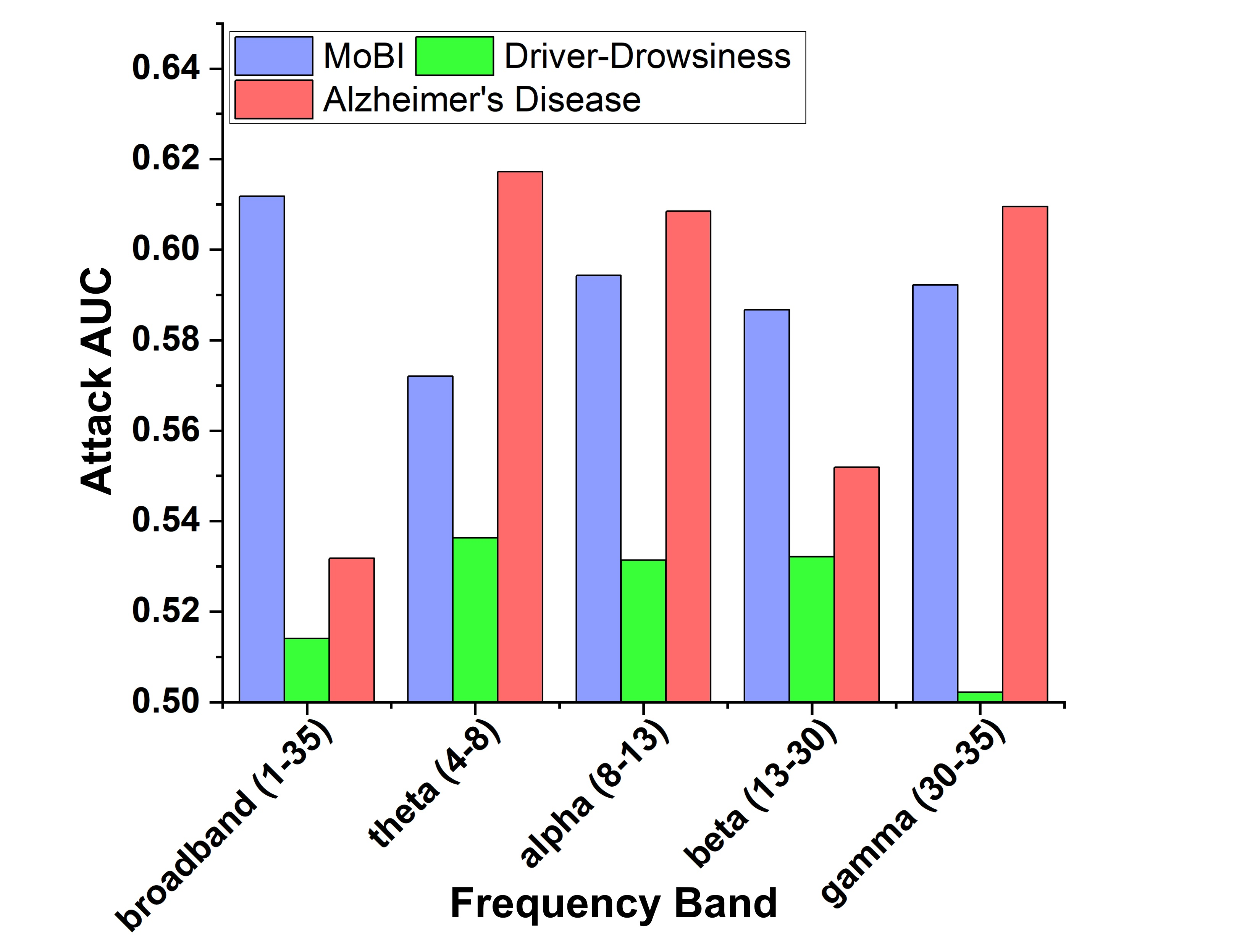, width=.9\columnwidth}}
  \caption{\textbf{Frequency bands.} Attack AUC for EEG inputs band-limited to theta, alpha, beta, or gamma, compared with the full broadband signal.}
  \vspace{-4pt}
  \label{fig:freqbandvsauc}
\end{figure}

\textbf{Frequency Bands:} Next, we restrict the input to individual frequency bands (Figure \ref{fig:freqbandvsauc}). On the MoBI dataset, theta, alpha, beta, and gamma produce nearly identical attack performance (Kruskal-Wallis \(p=0.84\); AUC spread = 0.005), with each individual band yielding an AUC of approximately 0.59, compared with 0.612 for the broadband signal (all \(p<0.001\)). Importantly, all individual bands still produce attack performance significantly above random guess. The same pattern is not fully consistent across the additional datasets: the Driver-Drowsiness dataset shows little difference between bands, with AUCs remaining close to 0.5, whereas in the Alzheimer’s Disease dataset several individual bands achieve AUCs comparable to or higher than the broadband condition. Overall, these results suggest that no specific frequency band consistently predicts MIA vulnerability across datasets. 

\begin{figure}[t]
  \centering
   {\epsfig{file = 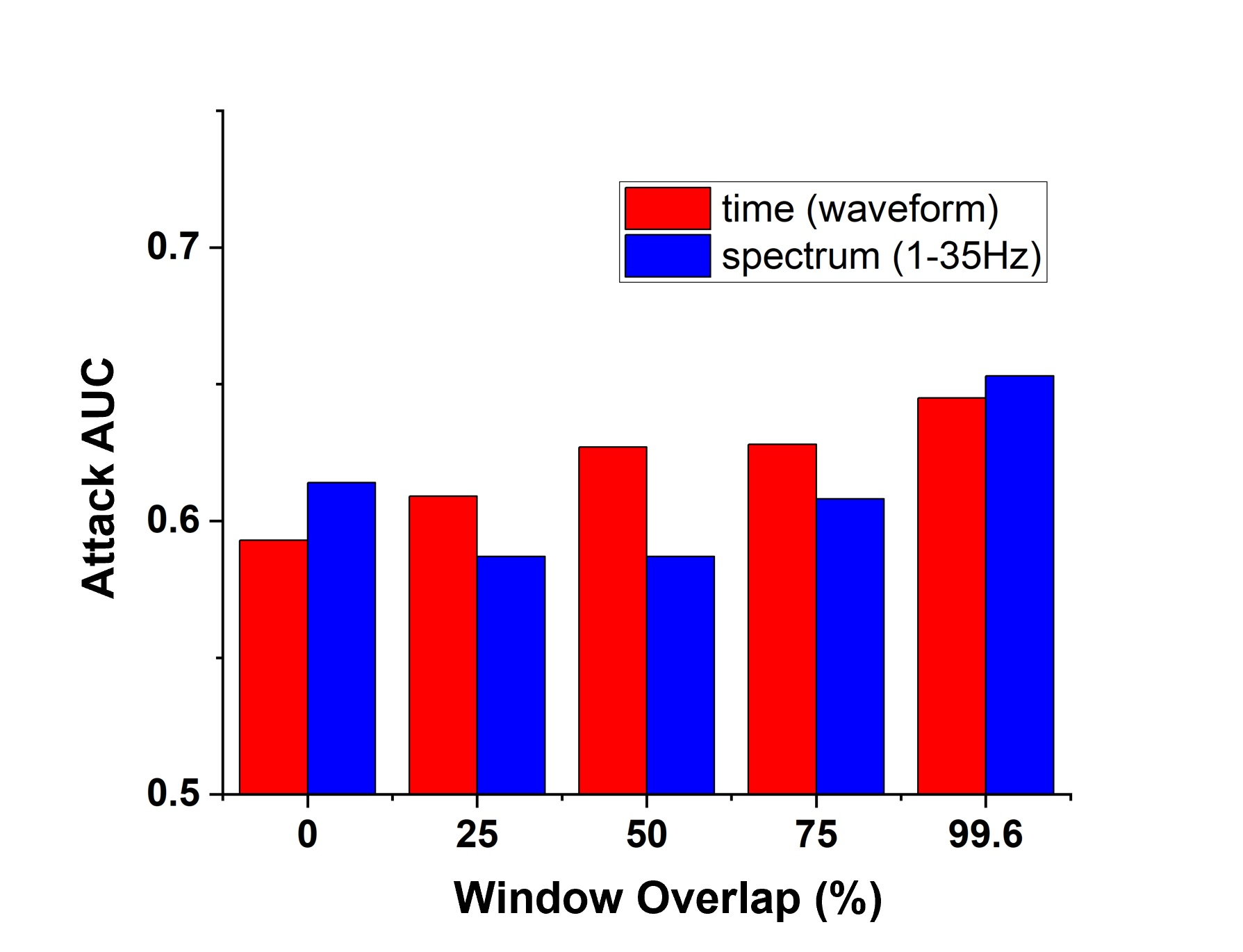, width=.9\columnwidth}}
  \caption{\textbf{Spectral representation.} Attack AUC for the time-domain waveform (red) versus the magnitude-spectrum (1–35 Hz, blue) input representation on the MoBI dataset. The two representations yield comparable attack strength at every overlap, with neither consistently superior.}
  \label{fig:Spectimeauc}
\end{figure}

\textbf{Input Representation:} Replacing the time-domain EEG waveform with its 1-35 Hz magnitude spectrum also has little effect on vulnerability (Figure \ref{fig:Spectimeauc}). Across the five MoBI window overlap conditions, the two representations yield comparable attack performance: mean AUC is 0.621 for the waveform and 0.610 for the spectral representation, with no significant difference between them (\(p=0.40\)). The spectrum performs marginally better at non-overlap and 1-stride, whereas the waveform performs marginally better at intermediate overlaps. Restricting the spectral representation to an individual frequency band again reduces attack performance relative to the full spectrum (\(p=0.0002\)). These findings indicate that membership information is accessible from both temporal and spectral representations rather than being tied to a particular input domain. 

\begin{figure}[t]
  \centering
   {\epsfig{file = 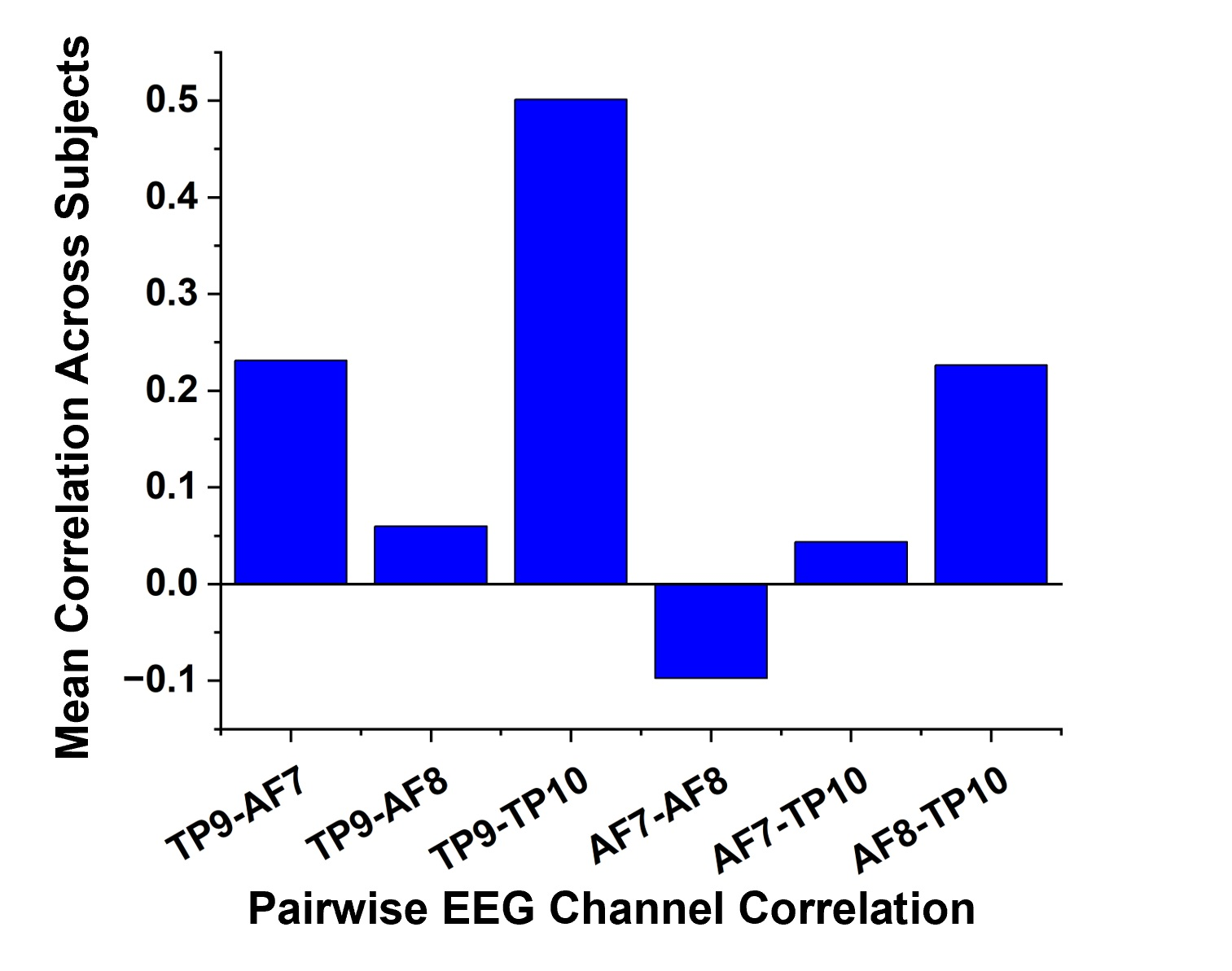, width=.9\columnwidth}}
  \caption{\textbf{Inter-channel correlation.} Mean correlation vs. channel pairs for MoBI dataset.}
  \label{fig:corr}
\end{figure}

\textbf{Inter-Channel Correlation:} Finally, we examine whether inter-channel correlation is associated with MIA vulnerability. The results are shown in Figure \ref{fig:corr}. At single-sample stride (1-stride), the average inter-channel correlation shows only a weak and non-significant relationship with attack performance ($r=0.16$, $p=0.65$). MIAs using the six pairwise correlation values achieve an AUC of approximately 0.54 at higher window overlaps, indicating that correlation structure contains some membership-related information but is a weak predictor. The strongest correlation is observed between the two temporal electrodes, TP9-TP10, as shown in Figure \ref{fig:corr}.

\subsection{Choice of MIA Method}

As explained in Section \ref{sec:MIAattacks}, there exist multiple methods with which a MIA can be executed, e.g., loss, confidence, modified entropy, and LiRA attack methods. Our results assumed the loss attack by default, consistent with \cite{cobilean2025investigating}. In this section, we explore the impacts of different attacks.

\begin{figure}[t]
  \centering
   {\epsfig{file = 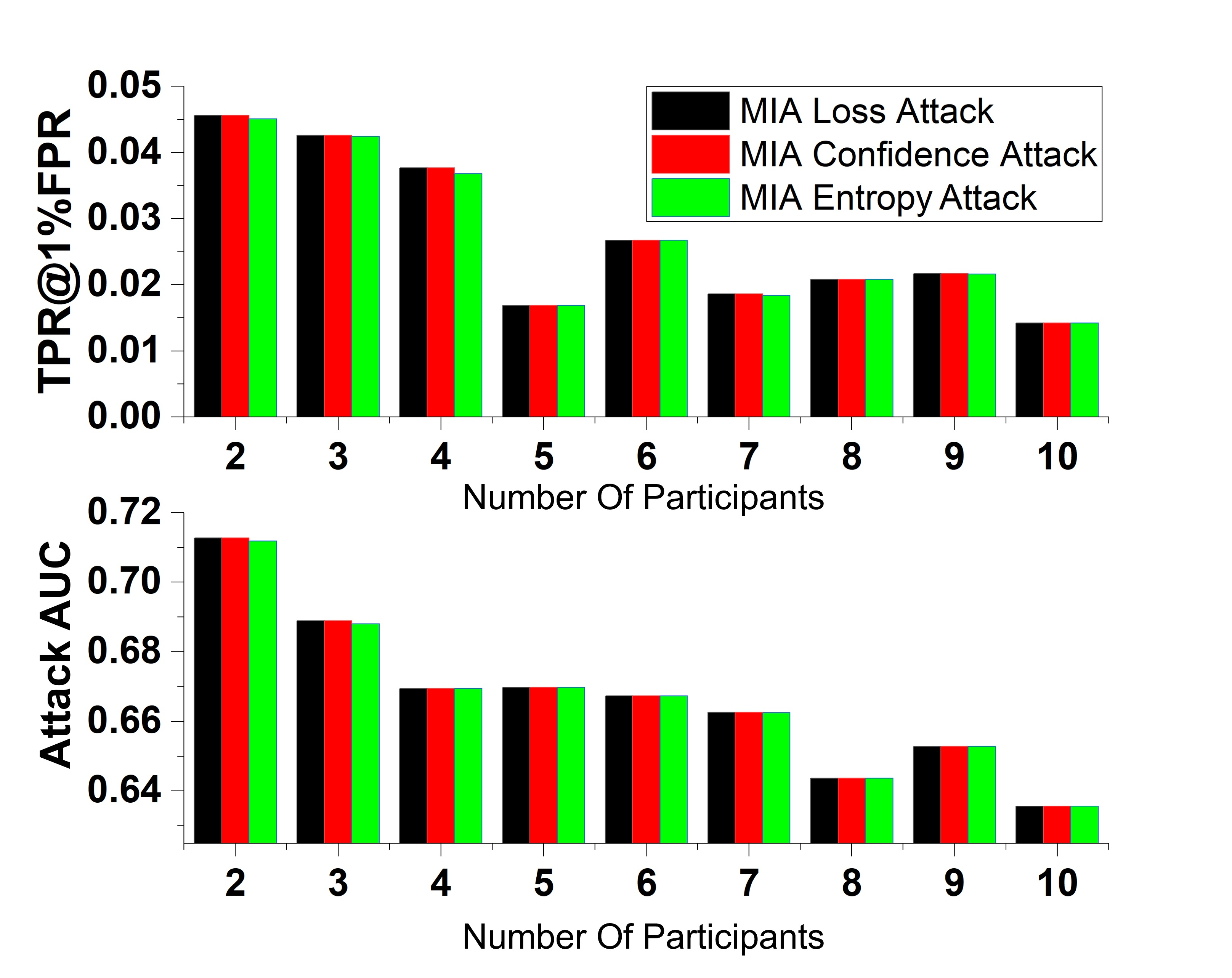, width=.9\columnwidth}}
  \caption{AUC and TPR @ 1\% FPR results of the loss, confidence, and modified entropy attacks on the MoBI dataset.}
  \label{fig:equivalenceattack}
\end{figure}

To confirm that our findings and conclusions do not depend on the specific choice of attack method, we first compare the loss, confidence, and modified entropy attacks. We performed this comparison across all six model architectures, three datasets, and multiple window overlap settings of the MoBI dataset. In general, we observed that the three attacks yield very similar AUCs (maximum difference of 0.02). Due to the page limit, we do not report all results in the paper, but only include Figure \ref{fig:equivalenceattack} as a representative result. Considering the highly similar performances of the three attack methods, we conclude that the loss attack is a good representative and suitable to adopt as our primary attack throughout our paper. We note that this high similarity between the attacks is caused by the fact that all datasets perform binary classification, i.e., the number of classes is $K$ = 2. Attacks can perform quite differently under multiclass classification. 

\begin{figure}[t]
  \centering
   {\epsfig{file = 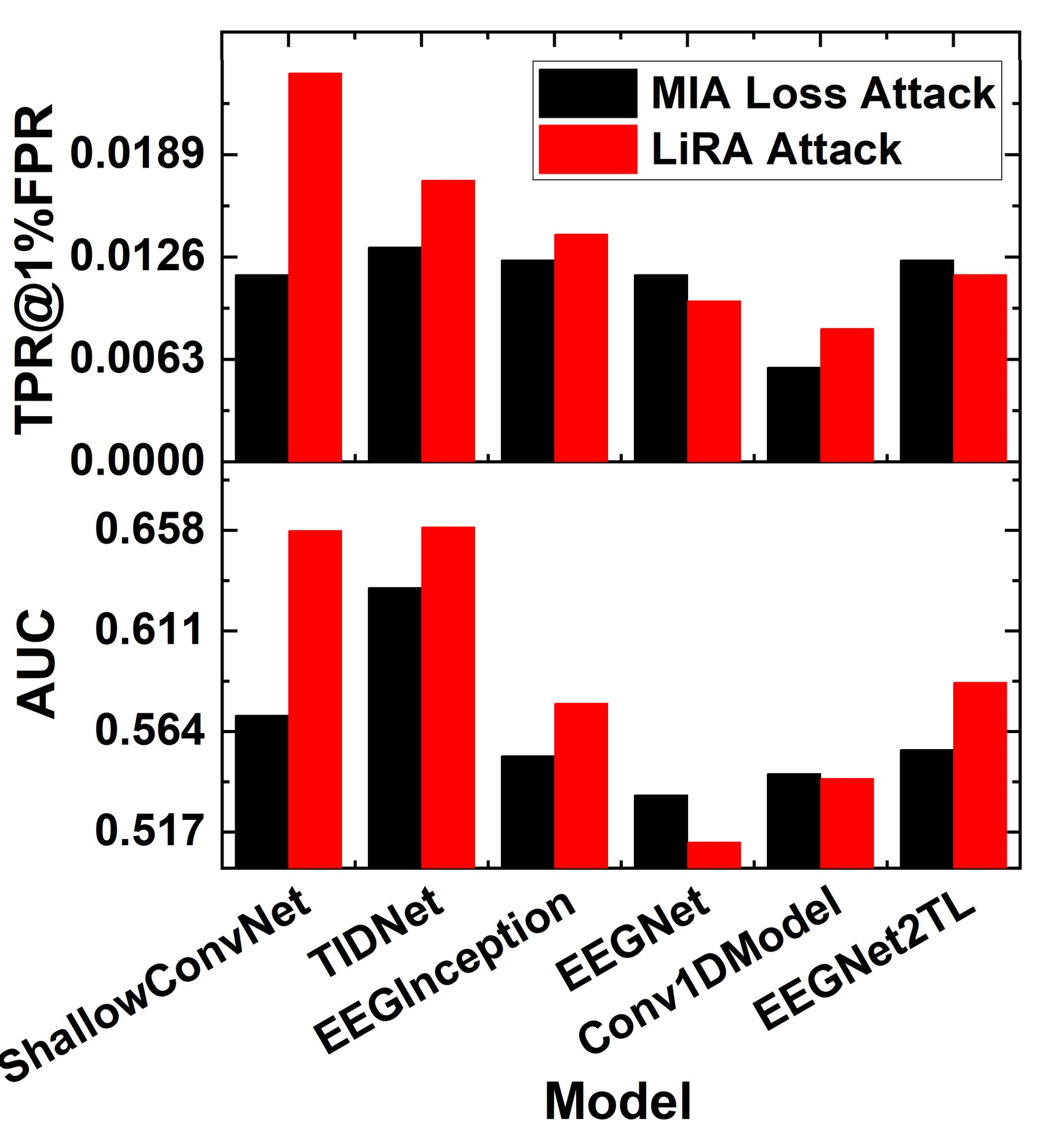, width=.7\columnwidth}}
  \caption{Comparison between LiRA and loss attack AUC and TPR @ 1\% FPR on the MoBI dataset.}
  \label{fig:liraloss}
\end{figure}

Another point of consideration is how our LiRA attack performs compared to the loss attack. We report the results of this experiment in Figure \ref{fig:liraloss}. We observe that LiRA generally outperforms the loss attack, as mean AUC rises from 0.570 to 0.606 and TPR @ 1\% FPR increases by roughly 3x. The improvement is largest for the more vulnerable architectures and negligible for the least vulnerable (e.g.: EEGNet), indicating that the choice of MIA method amplifies existing MIA vulnerability. Nevertheless, a practical consequence of this result is that focusing only on the loss attack may underestimate worst-case MIA risks. Proposing new, better-performing MIAs would be a valuable direction for future work, as there seems to be room for improvement in attack effectiveness as the attack method is changed.

\subsection{Can Subjects Be Identified?} \label{sec:subjID}

\begin{figure}[t]
  \centering
   {\epsfig{file = 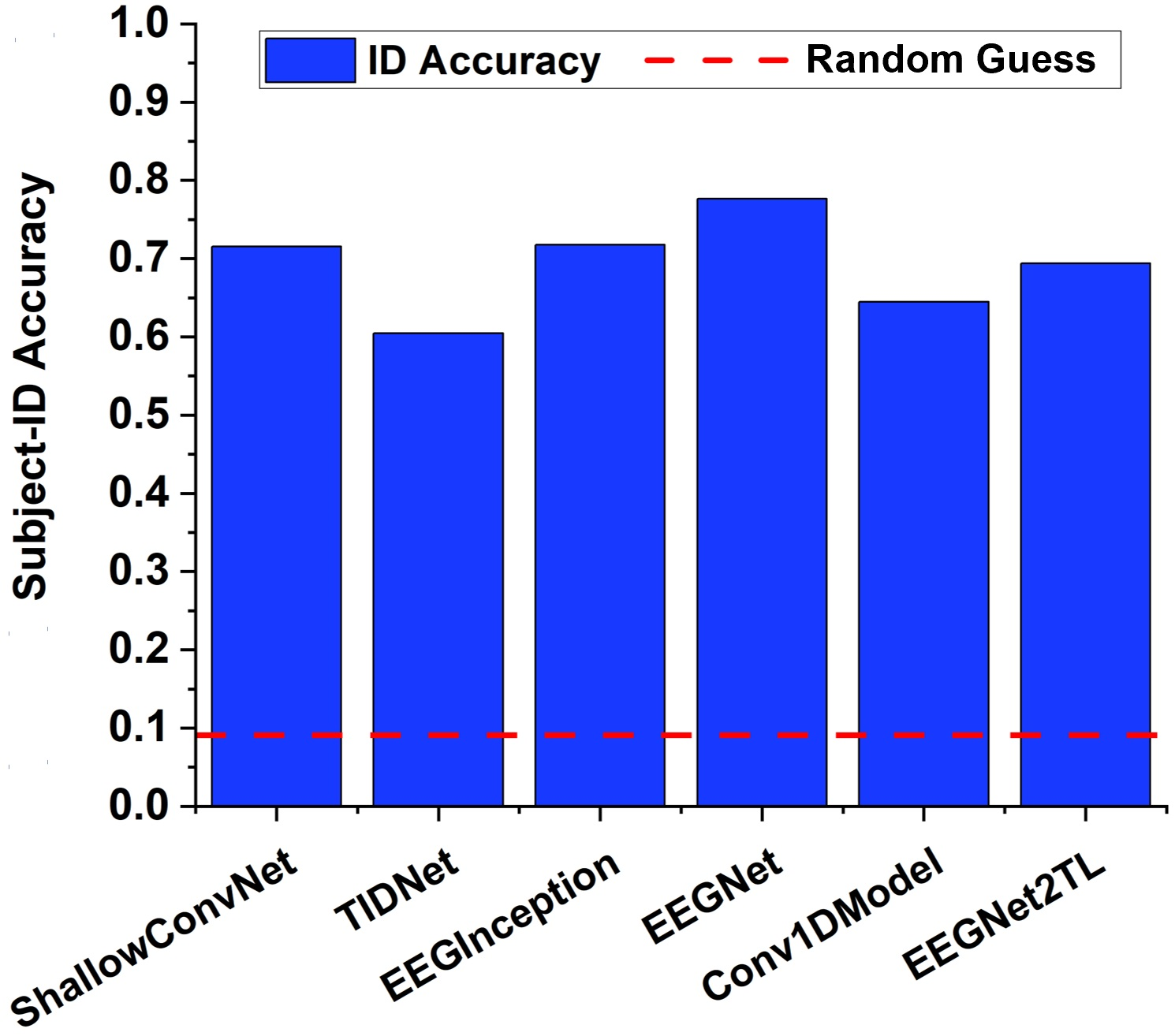, width=.8\columnwidth}}
  \caption{Results of the \textbf{subject identification accuracy} experiment on the MoBI dataset.}
  \label{fig:subid}
    \vspace{12pt}
\end{figure}

In order to check the relationship between MIA vulnerability and subject-specificity of EEG signals, we test whether subjects (participants in the datasets) can be identified using their EEG signal. We train our CNN model architectures by replacing the last decision layer with an N-way classification head to predict which subject (i.e., which participant) the given EEG window belongs to. The subject identification accuracy results of our models are given in Figure \ref{fig:subid}. Considering that our MoBI dataset contains 11 participants, the subject identification accuracy of random guessing is near 0.091. It can be observed from Figure \ref{fig:subid} that all architectures significantly outperform random guess, often achieving 6-7x accuracy of random guess (EEGNet particularly performs well). Based on these results, we conclude that EEG windows indeed carry subject-specific features, and participants can be identified using their EEGs. 

\begin{figure}[t]
  \centering
   {\epsfig{file = 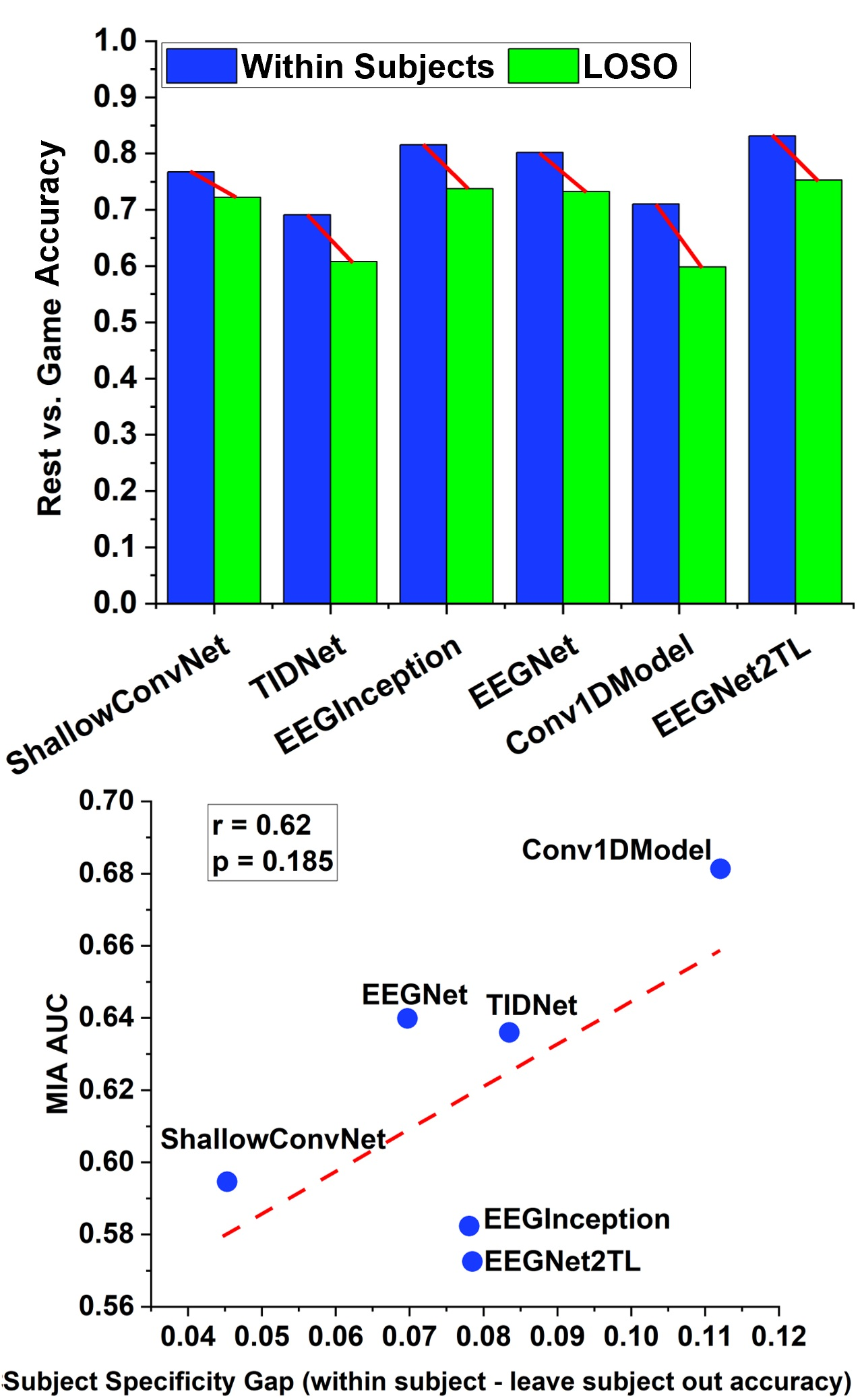, width=.8\columnwidth}}
  \caption{Accuracy on the rest vs game classification task (MoBI dataset) under within-subject and LOSO training settings.}
  \label{fig:val}
  \vspace{4pt}
\end{figure}

Finally, we revert to the original rest vs game classification task and compare models' accuracy under two settings: (i) when the same subjects' data is used in both $\mathcal{D}_{\mathrm{train}}$ and $\mathcal{D}_{\mathrm{test}}$, and (ii) under leave-one-subject-out (LOSO) training where the same subject's data cannot exist simultaneously in $\mathcal{D}_{\mathrm{train}}$ and $\mathcal{D}_{\mathrm{test}}$. The results are shown in Figure \ref{fig:val}. We observe that rest vs game task accuracy drops under the LOSO setting compared to the first setting. This shows that models perform better on the test data if they see data from the same subjects during training, reinforcing our prior subject-specificity finding. Furthermore, the second plot in Figure \ref{fig:val} shows the positive correlation between MIA AUC and the difference between within-subject accuracy and LOSO accuracy (though this correlation is not statistically significant). This implies that the models that generalize less to unseen subjects (i.e., the difference between within-subject and LOSO accuracy is higher) are more vulnerable to MIAs (i.e., MIA AUC is higher). Taken together with the prior results, these findings suggest that reliance on subject-specificity of EEG is a driving factor of membership leakage, as models that benefit most from seeing the same subjects during training also exhibit greater MIA vulnerability.

\section{\uppercase{Conclusion}}
\label{sec:conclusion}

In this paper, we investigated the factors influencing MIA vulnerability in EEG models across multiple datasets, CNN architectures, and MIA methods. Our results showed that poor model generalization (i.e., large train-test loss and accuracy gaps) is strongly associated with greater MIA vulnerability. Increasing the number of participants generally reduced attack success, whereas the amount of EEG data contributed per participant had a comparatively weaker effect. Model capacity and training duration also increased vulnerability, while no particular scalp region, frequency band, or input representation consistently explained membership leakage. Moreover, our subject identification and LOSO experiments indicated that reliance on subject-specific EEG representations contributes to MIA vulnerability. 

There exist multiple directions for future work. First, our attack performs MIA on EEG windows (i.e., segments); future work can also consider subject-level attacks. Second, we train CNN models with different architectures; future work can study EEG foundation models. Finally, defending against MIAs in the EEG context is an important avenue for future work. 

\vspace{-12pt}
\section*{Acknowledgments}
\vspace{-2pt}

This research was supported by the Scientific and Technological Research Council of Türkiye (TÜBİ\-TAK) under project number 123C595, the KUIS AI Thematic Seed Research Fund, and the BAGEP Outstanding Young Scientist Award. The authors thank TÜBİTAK, KUIS AI, and the Science Academy for their support.


\textbf{AI Disclosure:} We used Claude (Anthropic) as an AI assistant to revise the text and assist in debugging and documenting the Python code used in our experiments. All experiments were designed and executed by the authors, and all reported results, figures, and their interpretation are the authors' own. The authors take full responsibility for the paper contents.

\vspace{-10pt}

\bibliographystyle{apalike}
{\small
\bibliography{example}}

@article{liu2025recent,
  title={Recent applications of EEG-based brain-computer-interface in the medical field},
  author={Liu, Xiu-Yun and Wang, Wen-Long and Liu, Miao and Chen, Ming-Yi and Pereira, T{\^a}nia and Doda, Desta Yakob and Ke, Yu-Feng and Wang, Shou-Yan and Wen, Dong and Tong, Xiao-Guang and others},
  journal={Military Medical Research},
  volume={12},
  number={1},
  pages={14},
  year={2025},
  publisher={Springer}
}

@article{kiranyaz20211d,
  title={1D convolutional neural networks and applications: A survey},
  author={Kiranyaz, Serkan and Avci, Onur and Abdeljaber, Osama and Ince, Turker and Gabbouj, Moncef and Inman, Daniel J},
  journal={Mechanical Systems and Signal Processing},
  volume={151},
  pages={107398},
  year={2021},
  publisher={Elsevier}
}

@article{portillo2021mind,
  title={Mind the gap: State-of-the-art technologies and applications for EEG-based brain--computer interfaces},
  author={Portillo-Lara, Roberto and Tahirbegi, Bogachan and Chapman, Christopher AR and Goding, Josef A and Green, Rylie A},
  journal={APL Bioengineering},
  volume={5},
  number={3},
  year={2021},
  publisher={AIP Publishing}
}

@article{jwa2022addressing,
  title={Addressing privacy risk in neuroscience data: from data protection to harm prevention},
  author={Jwa, Anita S and Poldrack, Russell A},
  journal={Journal of Law and the Biosciences},
  volume={9},
  number={2},
  year={2022},
  publisher={Oxford University Press}
}

@inproceedings{meng2024protecting,
  title={Protecting multiple types of privacy simultaneously in EEG-based brain-computer interfaces},
  author={Meng, Lubin and Jiang, Xue and Jia, Tianwang and Wu, Dongrui},
  booktitle={2024 IEEE International Conference on Systems, Man, and Cybernetics (SMC)},
  pages={2894--2899},
  year={2024},
  organization={IEEE}
}

@inproceedings{mandal2022sok,
  title={SoK: Your mind tells a lot about you: On the privacy leakage via brainwave devices},
  author={Mandal, Anuradha and Saxena, Nitesh},
  booktitle={Proceedings of the 15th ACM Conference on Security and Privacy in Wireless and Mobile Networks},
  pages={175--187},
  year={2022}
}

@inproceedings{shokri2017membership,
  title={Membership inference attacks against machine learning models},
  author={Shokri, Reza and Stronati, Marco and Song, Congzheng and Shmatikov, Vitaly},
  booktitle={2017 IEEE Symposium on Security and Privacy (SP)},
  pages={3--18},
  year={2017},
  organization={IEEE}
}

@article{cobilean2025investigating,
  title={Investigating membership inference attacks against CNN models for BCI systems},
  author={Cobilean, Victor and Mavikumbure, Harindra S and Drake, Devin and Stuart, Morgan and Manic, Milos},
  journal={IEEE Journal of Biomedical and Health Informatics},
  year={2025},
  publisher={IEEE}
}

@inproceedings{yeom2018privacy,
  title={Privacy risk in machine learning: Analyzing the connection to overfitting},
  author={Yeom, Samuel and Giacomelli, Irene and Fredrikson, Matt and Jha, Somesh},
  booktitle={2018 IEEE 31st Computer Security Foundations Symposium (CSF)},
  pages={268--282},
  year={2018},
  organization={IEEE}
}

@inproceedings{song2021systematic,
  title={Systematic evaluation of privacy risks of machine learning models},
  author={Song, Liwei and Mittal, Prateek},
  booktitle={30th USENIX Security Symposium (USENIX Security 21)},
  pages={2615--2632},
  year={2021}
}

@inproceedings{carlini2022membership,
  title={Membership inference attacks from first principles},
  author={Carlini, Nicholas and Chien, Steve and Nasr, Milad and Song, Shuang and Terzis, Andreas and Tramer, Florian},
  booktitle={2022 IEEE Symposium on Security and Privacy (SP)},
  pages={1897--1914},
  year={2022},
  organization={IEEE}
}

@article{campisi2014brain,
  title={Brain waves for automatic biometric-based user recognition},
  author={Campisi, Patrizio and La Rocca, Daria},
  journal={IEEE Transactions on Information Forensics and Security},
  volume={9},
  number={5},
  pages={782--800},
  year={2014},
  publisher={IEEE}
}

@article{demuru2020eeg,
  title={EEG fingerprinting: Subject-specific signature based on the aperiodic component of power spectrum},
  author={Demuru, Matteo and Fraschini, Matteo},
  journal={Computers in Biology and Medicine},
  volume={120},
  pages={103748},
  year={2020},
  publisher={Elsevier}
}

@article{ravindran2017multi,
  title={Multi-modal mobile brain-body imaging (MoBI) dataset for assaying neural and head movement responses associated with creative video game play in children},
  author={Ravindran, Akshay Sujatha and Cruz-Garza, JG and Kopteva, A and Paek, A and Mobiny, A and Hernandez, Z and Contreras-Vidal, JL},
  journal={IEEE Dataport},
  year={2017}
}

@article{hu2022membership,
  title={Membership inference attacks on machine learning: A survey},
  author={Hu, Hongsheng and Salcic, Zoran and Sun, Lichao and Dobbie, Gillian and Yu, Philip S and Zhang, Xuyun},
  journal={ACM Computing Surveys (CSUR)},
  volume={54},
  number={11s},
  pages={1--37},
  year={2022},
  publisher={ACM New York, NY}
}

@article{yang2025regulating,
  title={Regulating neural data processing in the age of BCIs: ethical concerns and legal approaches},
  author={Yang, Hong and Jiang, Li},
  journal={Digital Health},
  volume={11},
  year={2025},
  publisher={SAGE Publications Sage UK: London, England}
}

@article{khanam2026evaluating,
  title={Evaluating multi-level membership inference risk in federated EEG learning},
  author={Khanam, Taslima and Siuly, Siuly and Wang, Kate and Whittaker, Frank and Wang, Hua},
  journal={Brain Informatics},
  year={2026},
  publisher={Springer}
}

@article{bai2024membership,
  title={Membership inference attacks and defenses in federated learning: A survey},
  author={Bai, Li and Hu, Haibo and Ye, Qingqing and Li, Haoyang and Wang, Leixia and Xu, Jianliang},
  journal={ACM Computing Surveys (CSUR)},
  volume={57},
  number={4},
  pages={1--35},
  year={2024},
  publisher={ACM New York, NY}
}

@article{sujatha2019assaying,
  title={Assaying neural activity of children during video game play in public spaces: a deep learning approach},
  author={Ravindran, Akshay Sujatha and Mobiny, Aryan and Cruz-Garza, Jesus G and Paek, Andrew and Kopteva, Anastasiya and Contreras Vidal, Jos{\'e} L},
  journal={Journal of Neural Engineering},
  volume={16},
  number={3},
  pages={036028},
  year={2019},
  publisher={IOP Publishing}
}

@article{mullen2015real,
  title={Real-time neuroimaging and cognitive monitoring using wearable dry EEG},
  author={Mullen, Tim R and Kothe, Christian AE and Chi, Yu Mike and Ojeda, Alejandro and Kerth, Trevor and Makeig, Scott and Jung, Tzyy-Ping and Cauwenberghs, Gert},
  journal={IEEE transactions on Biomedical Engineering},
  volume={62},
  number={11},
  pages={2553--2567},
  year={2015},
  publisher={IEEE}
}

@article{schirrmeister2017deep,
  title={Deep learning with convolutional neural networks for EEG decoding and visualization},
  author={Schirrmeister, Robin Tibor and Springenberg, Jost Tobias and Fiederer, Lukas Dominique Josef and Glasstetter, Martin and Eggensperger, Katharina and Tangermann, Michael and Hutter, Frank and Burgard, Wolfram and Ball, Tonio},
  journal={Human Brain Mapping},
  volume={38},
  number={11},
  pages={5391--5420},
  year={2017},
  publisher={Wiley Online Library}
}

@article{lawhern2018eegnet,
  title={EEGNet: a compact convolutional neural network for EEG-based brain--computer interfaces},
  author={Lawhern, Vernon J and Solon, Amelia J and Waytowich, Nicholas R and Gordon, Stephen M and Hung, Chou P and Lance, Brent J},
  journal={Journal of Neural Engineering},
  volume={15},
  number={5},
  pages={056013},
  year={2018},
  publisher={iOP Publishing}
}

@article{santamaria2020eeg,
  title={EEG-inception: a novel deep convolutional neural network for assistive ERP-based brain-computer interfaces},
  author={Santamaria-Vazquez, Eduardo and Martinez-Cagigal, Victor and Vaquerizo-Villar, Fernando and Hornero, Roberto},
  journal={IEEE Transactions on Neural Systems and Rehabilitation Engineering},
  volume={28},
  number={12},
  pages={2773--2782},
  year={2020},
  publisher={IEEE}
}

@article{kostas2020thinker,
  title={Thinker invariance: enabling deep neural networks for BCI across more people},
  author={Kostas, Demetres and Rudzicz, Frank},
  journal={Journal of Neural Engineering},
  volume={17},
  number={5},
  pages={056008},
  year={2020},
  publisher={IOP Publishing}
}

@article{miltiadous2023dataset,
  title={A dataset of scalp EEG recordings of Alzheimer’s disease, frontotemporal dementia and healthy subjects from routine EEG},
  author={Miltiadous, Andreas and Tzimourta, Katerina D and Afrantou, Theodora and Ioannidis, Panagiotis and Grigoriadis, Nikolaos and Tsalikakis, Dimitrios G and Angelidis, Pantelis and Tsipouras, Markos G and Glavas, Euripidis and Giannakeas, Nikolaos and others},
  journal={Data},
  volume={8},
  number={6},
  pages={95},
  year={2023},
  publisher={MDPI}
}

@article{cui2022compact,
  title={A compact and interpretable convolutional neural network for cross-subject driver drowsiness detection from single-channel EEG},
  author={Cui, Jian and Lan, Zirui and Liu, Yisi and Li, Ruilin and Li, Fan and Sourina, Olga and M{\"u}ller-Wittig, Wolfgang},
  journal={Methods},
  volume={202},
  pages={173--184},
  year={2022},
  publisher={Elsevier}
}

@article{cao2019multi,
  title={Multi-channel EEG recordings during a sustained-attention driving task},
  author={Cao, Zehong and Chuang, Chun-Hsiang and King, Jung-Kai and Lin, Chin-Teng},
  journal={Scientific Data},
  volume={6},
  number={1},
  pages={19},
  year={2019},
  publisher={Nature Publishing Group UK London}
}
\end{document}